\documentclass[secnumarabic,aps,prd,floatfix,nofootinbib,showkeys,twocolumn,showpacs,10pt,superscriptaddress] 
{revtex4-1}

\usepackage{times}
\usepackage{amsfonts}
\usepackage{amsmath}
\usepackage{amssymb}
\usepackage{natbib}
\usepackage{graphicx}
\usepackage{enumitem}
\usepackage{xcolor}
\usepackage[colorlinks=true,linkcolor=red, citecolor=blue]{hyperref}
\usepackage[outdir=./]{epstopdf}
\usepackage[normalem]{ulem}
\usepackage{amsmath}

\def\aap{A\&A}

\def\apj{ApJ}
\def\apjl{ApJL}

\def\jcap{JCAP}
\def\mnras{MNRAS}
\def\na{New Astronomy}

\def\prd{Phys. Rev. D}

\begin{document}
\definecolor{orange}{rgb}{0.9,0.45,0}
\def\CovDev{D}
\def\Res{{\mathcal R}}
\def\Gammaflat{\hat \Gamma}
\def\metricflat{\hat \gamma}
\def\Dflat{\hat {\mathcal D}}
\def\part_n{\partial_\perp}
%
\def\Lie{\mathcal{L}}
\def\A{\mathcal{X}}
\def\Aphi{\A_{\phi}}
\def\hAphi{\hat{\A}_{\phi}}
\def\E{\mathcal{E}}
\def\Ham{\mathcal{H}}
\def\M{\mathcal{M}}
\def\R{\mathcal{R}}
\def\p{\partial}
\def\hg{\hat{\gamma}}
\def\hA{\hat{A}}
\def\hD{\hat{D}}
\def\hE{\hat{E}}
\def\hR{\hat{R}}
\def\hcA{\hat{\mathcal{A}}}
\def\hDelt{\hat{\triangle}}
\def\na{\nabla}
\def\dif{{\rm{d}}}
\def\non{\nonumber}
\newcommand{\erf}{\textrm{erf}}
\newcommand{\saeed}[1]{\textcolor{blue}{SF: #1}} 
%
\renewcommand{\t}{\times}
\long\def\symbolfootnote[#1]#2{\begingroup%
\def\thefootnote{\fnsymbol{footnote}}\footnote[#1]{#2}\endgroup}
\title{Gravitational Lensing in More Realistic Dark Matter Halo Models} 

\author{Ali Tizfahm}
\email{a.tizfahm@email.kntu.ac.ir}
\affiliation{Department of Physics, K.N. Toosi University of Technology, P.O. Box 15875-4416, Tehran, Iran}
\affiliation{PDAT Laboratory, Department of Physics, K.N. Toosi University of Technology, P.O. Box 15875-4416, Tehran, Iran}

\author{Saeed Fakhry}
\email{s\_fakhry@sbu.ac.ir}
\affiliation{Department of Physics, Shahid Beheshti University, 1983969411, Tehran, Iran}
\affiliation{PDAT Laboratory, Department of Physics, K.N. Toosi University of Technology, P.O. Box 15875-4416, Tehran, Iran}

\author{Javad T. Firouzjaee} 
\email{firouzjaee@kntu.ac.ir}
\affiliation{Department of Physics, K.N. Toosi University of Technology, P.O. Box 15875-4416, Tehran, Iran}
\affiliation{PDAT Laboratory, Department of Physics, K.N. Toosi University of Technology, P.O. Box 15875-4416, Tehran, Iran}
\affiliation{School of Physics, Institute for Research in Fundamental Sciences (IPM), P.O. Box 19395-5531, Tehran, Iran}

\author{Antonino Del Popolo}
\email{antonino.delpopolo@unict.it}
\affiliation{Dipartimento di Fisica e Astronomia, University of Catania, Viale Andrea Doria 6, 95125 Catania, Italy}
\affiliation{INFN Sezione di Catania, Via Santa Sofia,64, 95123 Catania, Italy}

\date{\today}

\begin{abstract} 
\noindent
In this study, we explore gravitational lensing using more realistic dark matter halo models, moving beyond the limitations of spherical-collapse approximations. Through analytical calculations employing various mass functions, we address critical factors often neglected in the standard Press-Schechter formalism, such as ellipsoidal-collapse conditions, angular momentum dynamics, dynamical friction, and the cosmological constant. Our analysis incorporates two widely recognized halo density profiles, the Navarro-Frenk-White and Einasto profiles considering both spherical and ellipsoidal-collapse scenarios. We provide detailed calculations of key gravitational lensing observables, including Einstein radii, lensing optical depths, and time delays, across a broad range of redshifts and masses using two different lensing models: the point mass and singular isothermal sphere (SIS) models. Our results show that using more realistic dark matter halo models enhances lensing effects compared to their spherical-collapse counterparts. Additionally, our analyses of lensing optical depths and time delays reveal distinct differences between the point mass and SIS lens models. These findings underscore the importance of using realistic halo descriptions instead of simplified approximations when modeling gravitational lensing, as this approach can more accurately capture the complex structures of dark matter.
\end{abstract}

\keywords{Gravitational Lensing --- Dark Matter --- Halo Mass Function --- Optical Depth --- Ellipsoidal Collapse}

\maketitle
\vspace{0.8cm}

\section{Introduction} 
In the realm of gravitational lensing, the trajectory of null waves emmited from the source is altered when these waves traverse the vicinity of a massive compact object. This alteration in propagation direction is attributed to the gravitational influence of the massive object. The interference of these lensed waves gives rise to novel wave patterns. The formation of these patterns is contingent upon several factors, including the nature of the lens, the shape of the lensing, and the interference among different potential rays. Gravitational lensing is a universal phenomenon that affects all forms of radiation, including electromagnetic (EM) waves \citep{1998LRR.....1...12W, 2010CQGra..27w3001B, 2018GReGr..50...42C, 2023FrP....1113909C} and GWs \citep{2003ApJ...595.1039T, 2020PhRvD.101f4011H, 2021PhRvD.104j3529D, 2021PhRvD.103d4005H, 2022ApJ...926L..28B, 2023Univ....9..200G, 2023PhRvD.107j3023A, 2023PhRvD.108d3527T, 2023PhRvD.108j3529T, 2024CQGra..41a5005G}. This occurs when the path of wave is influenced by a massive celestial entity, such as a galaxy, a black hole, or a star. As postulated in Einstein's theory of general relativity, the curvature of spacetime, induced by a massive object, deviates the trajectory of null waves. This deviation engenders a distorted perception of the source and potentially leads to the observation of multiple images of the same source. 

The lensing of EM waves, in particular, has been instrumental in propelling research in the fields of astrophysics and cosmology. This underscores the pivotal role of lensing phenomena in our understanding of the Universe. Gravitational lensing of EM waves has facilitated significant advancements in astronomical research. For instance, it has enabled the discovery of exoplanets orbiting distant stars \cite{bond2004ogle}, revealed phase effects \cite{ezquiaga2021phase}, and allowed the observation of otherwise undetectable faint objects \cite{welch2022highly}. Furthermore, gravitational lensing has been instrumental in exploring the nature of dark matter \cite{massey2010dark} and evaluating cosmological parameters \cite{aghanim2020planck}. Analogous to the lensing of EM waves, GWs can also be lensed \cite{ohanian1974focusing}. In 2015, the gravitational wave event GW150914 was detected by the LIGO-Virgo detectors \cite{abbott2016observation}. This wave, produced by the merger of two distant black holes, marked a groundbreaking development in astrophysics, introducing a novel method for observing and understanding the Universe \cite{maggiore2007gravitational}.

Subsequently, new sources of merging black holes and neutron stars were detected by the LIGO-Virgo detectors \cite{2023gwtc}. Similar to EM lensing, the detection of lensed GWs has led to significant advancements in theoretical physics, astrophysics, and cosmology. Cosmography can be enhanced by utilizing the lensing effect on transient events, which are the most common type of GW event \cite{liao2022strongly}. Gravitational lensing also holds the potential to address the mass sheet degeneracy problem, which affects EM lensing \cite{2021breaking}. Additionally, GW birefringence \cite{ezquiaga2020gravitational} and the propagation speed of GWs \cite{baker2017multimessenger} can be used to test different theories of gravity. The microlensing effect on GW signals from black hole binaries, induced by low-mass dark matter halos that lack sufficient baryonic matter to form stars, is another area of interest \cite{2023JCAP...07..007F}. Given that everything in the Universe is subject to gravity, gravitational lensing offers a unique opportunity to study the dark aspects of cosmological structures.

One of the most perplexing aspects of cosmology pertains to the enigmatic nature of dark matter. Within this context, the gravitational interaction between null waves and dark matter substructures emerges as a promising avenue for potentially detecting dark matter using lensing observatories. In \cite{2022A&A...668A.166L}, broad generalizations are presented that emphasize the ability of sub-galactic dark matter halos to act as strong gravitational lenses for background compact sources, producing gravitational lensing events on milli-arcsecond scales, known as milli-lenses, across different dark halo matter models. Cold dark matter light halos, for instance, may serve as gravitational lenses for null waves passing nearby, with their abundance offering predictive insights into this phenomenon \cite{jung2019gravitational}. Conversely, the formation of dark matter halos has been investigated through numerical simulations and analytical methods, particularly focusing on those formed during the matter-dominated era \cite{klypin2011dark, 2018PhRvD..98f3527D, 2022MNRAS.517L..46W, 2023PDU....4101259D, 2024ARep...68...19D}. The formation and evolution of these halos necessitate well-defined models for halo mass functions that characterize the distribution of mass among dark matter halos. In the study of large-scale structures, significant research effort has been dedicated to identifying mass functions that accurately reflect the simulations and data pertaining to galactic halos \cite{press1974formation}.

It is well-established that the halo mass function stands as a pivotal quantity in the characterization of dark matter halos according to their mass. Specifically, the halo mass function delineates the mass distribution of structures whose overdensities have surpassed a predefined threshold, thereby undergoing gravitational collapse independent of the cosmic expansion. Consequently, in the pursuit of elucidating the enigmatic aspects of the Universe \cite{2021arXiv210700562G, 2023PDU....3901144G, 2023PhRvD.107f3507F, 2023arXiv231115307F, 2024PDU....4401477R}, one of the primary challenges encountered by cosmological models lies in furnishing a high-precision mass function for the distribution of dark matter \cite{2001MNRAS.323....1S, 2002MNRAS.329...61S, 2006ApJ...637...12D, 2017JCAP...03..032D}.

Specifically, as demonstrated in reference \cite{press1974formation}, Monte Carlo simulations are employed to derive the distribution of first barrier crossings for a Gaussian random walk, which subsequently informs the formulation of the Press-Schechter (PS) mass function. Press and Schechter introduced an analytical model incorporating primordial fluctuations that evolve from the linear phase to the collapse time, adhering to the spherical-collapse model. Despite its conceptual simplicity, the PS mass function demonstrates consistency with observational data gleaned from surveys. However, disparities arise between the PS mass function and numerical results at certain mass thresholds, underscoring its limitations in accurately describing the mass distribution of dark matter halos, particularly at high redshifts \cite{2006ApJ...637...12D, 2017JCAP...03..032D}. Such discrepancies may be attributed to the neglect of certain physical effects within the PS formalism, which are instrumental in predicting halo abundance.

In this context, Sheth and Tormen (ST) introduced a formalism rooted in a more realistic model, demonstrating improved concordance with simulation data \cite{2001MNRAS.323....1S, 2002MNRAS.329...61S}. The ST formalism is founded upon an ellipsoidal-collapse model, characterized by a dynamically evolving threshold value of overdensities. However, semi-analytical investigations indicate that the ST mass function tends to overestimate the number of halos at high masses \cite{reed2003evolution}. In light of observations from LIGO and Virgo, it becomes evident that the merger rate within each halo, as predicted by the more realistic halo models, exceeds that of the spherical models \cite{2021PhRvD.103l3014F, 2022PhRvD.105d3525F, 2022ApJ...941...36F, 2023PDU....4101244F, 2023ApJ...947...46F, 2023arXiv230811049F, 2024ApJ...966..235F, 2024arXiv240811995F}.

In this work, we focus on the gravitational lensing of null waves in more realistic dark matter halo models. In this respect, the outline of the work is as follows. In Sec. \ref{sec:ii} we introduce the dark matter halo models, including density profiles like NFW and Einasto as well as more realistic mass functions. In Sec. \ref{sec:iii} we explain our methodology for calculating gravitational lensing observables like Einstein radii, lensing optical depths, and lensing time delays for different halo models. We also perform calculations and present results on these lensing observables over a range of lens/source redshifts and masses, highlighting differences between spherical and ellipsoidal-collapse scenarios. In Sec. \ref{sec:iv} we discuss the implications of our results and potential directions of future work. It is important to mention that in this study, we have employed the geometrical optics approximation and have set the cosmological parameters based on \cite{aghanim2020planck}.

\section{Dark Matter Halo Models}\label{sec:ii}
According to the widely accepted cosmological model, dark matter halos are regarded as nonlinear cosmic structures that have been dispersed throughout the Universe due to the formation and evolution of hierarchical structures. The primordial density fluctuations could have surpassed a critical threshold, undergone gravitational collapse, and become qualified to form these dark matter halos. From a physical standpoint, these conditions can be characterized by a dimensionless quantity known as the density contrast, derived from the excursion sets theory. This quantity is expressed as $\delta(r) \equiv (\rho(r) - \bar{\rho})/\bar{\rho}$, where $\rho(r)$ represents the density of the overdense region at an arbitrary point $r$, and $\bar{\rho}$ denotes the mean density of the background.

Alternatively, the inner regions of dark matter haloes, which play a crucial role in shaping cosmological models and structure formation, can be characterized by a function dependent on radius, known as the density profile. Spectroscopic observations of gravitational lensing, X-ray temperature maps, and stellar dynamics in galaxies serve as reliable methods to estimate the distribution of dark matter in the core areas of galactic haloes. Over the past few decades, both analytical methods and numerical simulations have been employed to derive an appropriate density profile that aligns its predictions with the respective observational data. Within the context of cold dark matter models, Navarro, Frenk, and White (NFW) proposed a specific density profile based on the $N$-body simulations \citep{1996ApJ...462..563N}
\begin{equation}\label{nfw}
\rho(r)=\frac{\rho_{\rm s}}{r/r_{\rm s}(1+r/r_{\rm s})^2},
\end{equation}
where $\rho_{\rm s}$ and $r_{\rm s}$ represent the scale density and radius, respectively, which vary from halo to halo. Another profile that is used in dark matter simulations, widely known as the Einasto profile, possesses a power-law logarithmic slope of the form \citep{1965TrAlm...5...87E}
\begin{equation}\label{ein}
\rho(r) = \rho_{\rm s} \exp\left(-\frac{2}{\alpha}\left[\left(\frac{r}{r_{\rm s}}\right)^{\alpha}-1\right]\right),
\end{equation}
where $\alpha$ is the shape parameter. Note that for both of the above forms, one has
\begin{equation}
\frac{d\ln{\rho(\rm r)}}{d\ln{r}} = -2  \quad \text{for} \quad \frac{r}{r_{\rm s}}=1,
\end{equation}
i.e. the logarithmic slope of the density distribution is $-2$.

Besides the density profile, the central density of galactic halos can also be characterized by the concentration parameter, defined as follows
\begin{equation}
    C \equiv \frac{r_{\rm vir}}{r_{\rm s}},
\end{equation}

where $r_{\rm vir}$ represents the halo's virial radius. This radius encompasses a volume in which the average halo density is $200$ to $500$ times the critical density of the Universe. Both numerical simulations and analytical methods suggest that the concentration parameter should vary dynamically with mass and cosmological redshift for accurate predictions. Numerous studies have aimed to determine an appropriate concentration parameter \citep{2012MNRAS.423.3018P, 2014MNRAS.441.3359D, 2016MNRAS.456.3068O, 2016MNRAS.460.1214L}. In this study, we adopt Eq.\,(C1) from \citep{2016MNRAS.460.1214L} to represent the concentration-mass-redshift relation for spherical-collapse dark matter halo models. For more realistic dark matter halo models, we use Eqs.\,(34) and (36) from \citep{2016MNRAS.456.3068O}.

The mass contained within the virial radius of a halo, known as the virialized mass, can be determined by the following integral:
\begin{equation} 
M_{\rm vir} = \int_{0}^{r_{\text{vir}}} 4\pi r^2 \rho(r) dr.
\end{equation}
By substituting the density profiles given in Eqs.\,\eqref{nfw} and \eqref{ein} into the above relation and evaluating the integral, one can obtain the virialized mass for the NFW profile as:
\begin{equation}
M_{\rm vir (\rm NFW)} = 4\pi\rho_{s}r_{s}^{3}\left(\ln (1+C)-\frac{C}{1+C}\right),
\end{equation}
and for the Einasto profile as:
\begin{equation}
M_{\rm vir (\rm Ein)} = 4\pi\rho_{s}r_{s}^{3}l(C,\alpha).
\end{equation}
In the latter equation, $l(C,\alpha)$ is a function that depends on the concentration and shape parameter, with the following form:
\begin{equation}
l(C,\alpha)=\frac{\exp(2/\alpha)}{\alpha}\left(\frac{\alpha}{2}\right)^{3/\alpha}\Gamma\left(\frac{3}{\alpha},\frac{2}{\alpha}C^{\alpha}\right),
\end{equation}
where $\Gamma(x, y) = \int_{0}^{y} t^{x-1} e^{-t} dt$ is the incomplete Gamma function.

Dark matter halos offer a practical and fundamental means to study nonlinear gravitational collapse in the Universe. Understanding the mass distribution of these halos from a statistical perspective can enhance our comprehension of their underlying physics. The halo mass function is introduced to describe this mass distribution within a given volume. Essentially, the halo mass function quantifies the masses of structures that exceed a certain density threshold, are unaffected by the Universe's expansion, and are destined to collapse. As the Universe expands, density contrasts can surpass the linear regime, entering the nonlinear regime. At this point, overdense regions decouple from the expanding Universe, enter a turnaround phase, and collapse, resulting in the formation of structures.

In the standard model of cosmology, the halo mass function can be analytically derived using the excursion set theory, which treats the density field as a stochastic process across different scales. The foundation of excursion set theory is the spherical collapse model, which establishes the threshold overdensity necessary for the collapse of a spherical perturbation \citep{press1974formation}. For an Einstein de Sitter Universe and a collapsing spherical halo model, this threshold overdensity can be determined as
\begin{equation}
\delta_{\rm sc}=\frac{3(12\pi)^{2/3}}{20}\left(1-0.01231\log\left[1+\frac{\Omega_{\rm m}^{-1}-1}{(1+z)^{3}}\right]\right),
\end{equation}
where $\Omega_{\rm m}$ represents the density parameter of the matter content. Consequently, the threshold overdensity can be approximated as $1.686$ within a narrow redshift range. A crucial aspect of the current analysis is that $\delta_{\rm sc}$ is nearly unaffected by the halo mass, which potentially leads to some underestimations.

The initial refinement of the PS threshold was achieved in Ref.\,\citep{1998A&A...337...96D}, showing that the collapse threshold can also be mass dependent
\begin{equation}
\delta_{\rm cm}=\delta_{\rm sc}\left(1+\frac{\beta}{\nu^{\alpha}}\right).
\end{equation}
where $\alpha=0.585$ and $\beta=0.46$. Also, in Ref.\,\citep{2001MNRAS.323....1S}, the threshold is determined by examining an ellipsoidal collapse, outlined as follows:
\begin{equation}
\delta_{\rm ec}=\delta_{\rm sc}\left(1+\frac{\beta_1}{\nu^{\alpha_1}}\right).
\end{equation}
where $\alpha_1=0.615$ and $\beta_1=0.485$.

In this context, $\nu$ represents the peak height and is defined as
\begin{equation}
\nu(M,z)\equiv \frac{\delta_{\rm sc}}{\sigma(M, z)}=\frac{\delta_{\rm sc}}{D(z)\sigma(M, 0)},
\end{equation}
where $D(z)$ is the linear growth factor, and $\sigma(M, z)$ denotes the linear root-mean-square fluctuation of overdensities on a comoving scale including a mass $M$ at cosmological redshift $z$
\begin{equation}
\sigma^{2}(M,a)=\frac{1}{2\pi^{2}}\int_{0}^{\infty}P(k,a)W^{2}(k,M)k^{2}dk.
\end{equation}
In this relation, $P (k, a)$ refers to the power spectrum of linear matter fluctuations and $W (k, M)$ is the smoothing window function, assumed to be a sharp-$k$ filter
\begin{equation}
W(k, M) =
\left\{
\begin{array}{ll}
1 & \hspace*{0.5cm}\text{if } \,\, 0 < k \leq k_{M},\\
0 & \hspace*{0.5cm}\rm{otherwise},
\end{array}
\right.
\end{equation}
where $k_{M} = \left(6\pi^2 \rho_{\rm m,0}/M \right)^{1/3}$ and $\rho_{\rm m,0} = 0.33 \,{\rm M_{\odot} kpc^{-3}}$.

In the excursion set framework, the mass function is defined as the comoving number density of halos within a mass interval $(M, M + dM)$
\begin{equation}\label{massfunc}
n(M, z)=\frac{\rho_{\rm m,0}}{M^{2}} \left| \frac{{\rm d} \ln\nu}{{\rm d}\ln M} \right| \nu f(\nu),
\end{equation}
where $\nu f(\nu)$ is the multiplicity function and illustrates the distribution of the first crossing. The multiplicity function in the framework of PS formalism can be obtained as  follows \citep{press1974formation}
\begin{equation}
[\nu f(\nu)]_{\rm PS}=\sqrt{\frac{2}{\pi}}\frac{(\nu+0.556)\exp[-0.5(1+\nu^{1.34})^{2}]}{(1+0.0225\nu^{-2})^{0.15}}.
\end{equation}
However, there are differences between the predictions of the PS mass function and the distribution of dark matter halos. These discrepancies may be due to various physical factors that are not considered in the PS formalism but could significantly impact the abundance of dark matter halos. In this context, the ST mass function incorporates geometric corrections and extends the spherical collapse model of the PS formalism to the ellipsoidal collapse halo models. The ST mass function has the following form \citep{2001MNRAS.323....1S}
\begin{equation}
[\nu f(\nu)]_{\rm ST}=A_{1}\sqrt{\frac{2\nu^{\prime}}{\pi}}\left(1+\frac{1}{\nu^{\prime q}}\right)\exp\left(-\frac{\nu^{\prime}}{2}\right),
\end{equation}
where $q=0.3$, $\nu^\prime=0.707\nu^2$, and $A_{1}=0.322$, which is established by ensuring that the integral of $f(\nu)$ over all possible values of $\nu$ equals unity. 

In addition to geometric conditions during the virialization process of dark matter halos, other physical factors also influence the collapse of overdense regions and, consequently, the halo mass function. It is essential to include these factors in the analysis as they represent the true physics of halo collapse and growth, as well as the processes behind structure formation and evolution throughout cosmic history. This approach allows the collapse threshold to depend on effective physical factors, enabling the barrier to adjust accordingly and resulting in a more accurate model for halo collapse. Key corrections, such as the effects of angular momentum, dynamical friction, and the cosmological constant, are considered. These corrections help reduce discrepancies, particularly in debated mass ranges. When including the effects of angular momentum and the cosmological constant, the mass function, termed DP1 in this study, is found to be more realistic \citep{2006ApJ...637...12D}
\begin{equation}
[\nu f(\nu)]_{\rm DP1}=A_{2}\sqrt{\frac{\nu^{\prime}}{2\pi}}k(\nu^{\prime})\exp\left[-0.4019\nu^{\prime} l(\nu^{\prime})\right],
\end{equation}
where $A_{2}=0.974$ is determined by normalization, and
\begin{equation}
k(\nu^\prime)=\left(1+\frac{0.1218}{\nu^{\prime 0.585}}+\frac{0.0079}{\nu^{\prime 0.4}}\right),
\end{equation}
and 
\begin{equation}
l(\nu^\prime)=\left(1+\frac{0.5526}{\nu^{\prime 0.585}}+\frac{0.02}{\nu^{\prime 0.4}}\right)^2.
\end{equation}

The impact of dynamical friction on the barrier was examined in \citep{2017JCAP...03..032D}, resulting in the mass function referred to as DP2
\begin{equation}
[\nu f(\nu)]_{\rm DP2}=A_{3}\sqrt{\frac{\nu^\prime}{2\pi}}m(\nu^\prime)\exp[-0.305\nu^{\prime2.12} n(\nu^\prime)],
\end{equation}
where $A_{3}=0.937$ is set by normalization, and
\begin{equation}
m(\nu^\prime)=\left(1+\frac{0.1218}{\nu^{\prime 0.585}}+\frac{0.0079}{\nu^{\prime 0.4}}+\frac{0.1}{\nu^{\prime 0.45}}\right),
\end{equation}
and 
\begin{equation}\label{nnuprime}
n(\nu^\prime)=\left(1+\frac{0.5526}{\nu^{\prime 0.585}}+\frac{0.02}{\nu^{\prime 0.4}}+\frac{0.07}{\nu^{\prime 0.45}}\right)^2.
\end{equation}
In the subsequent sections, we will utilize these halo mass functions to analyze the gravitational lensing of GWs within more realistic dark matter halo models.

\section{Gravitational Lensing}\label{sec:iii}

Gravitational lensing of gravitational waves (GWs) offers unique opportunities that differ from those in the electromagnetic spectrum. It serves as a more direct probe of the gravitational potential and mass distribution of astrophysical lenses \citep{2023PhRvD.108d3527T}. The oscillatory nature of GWs allows wave optics effects, such as diffraction and interference, to occur in interesting ways during lensing. Observables, like time delays between multiple lensed images, can accurately encode information about lens masses and cosmological parameters without the ambiguities, such as the mass-sheet degeneracy, that affect strong lensing studies with light \citep{2013A&A...559A..37S}. Moreover, unlike distant quasars or galaxies typically used as light sources, GW sources such as compact binary mergers are extremely compact and bright, acting as essentially point-like sources, which simplifies lensing calculations. The transient nature of these events also allows for monitoring time variability in the lensing effects \citep{2018MNRAS.476.2220L}.

In order to get a rough estimate of the lens and source masses for which the interference effect is detectable, we start with the simplest case where the lens can be approximated as a point mass. Hence, the angular Einstein radius can be specified as \citep{2006JCAP...01..023M}, 
\begin{eqnarray}
\theta_{\rm EP}(M)=\sqrt{\dfrac{4 G M}{c^2}\dfrac{D_{\rm ls}}{D_{\rm l}D_{\rm s}}} \hspace*{3.3cm}\nonumber \\
\approx 3 \times 10^{-6}\left(\frac{M}{M_{\odot}}\right)^{1/2} \left[\frac{D_{\rm ls}/(D_{\rm l}D_{\rm s})}{1 {\rm Gpc}}\right]^{1/2} \rm arcsec,
\end{eqnarray}
where $D_{\rm s}$, $D_{\rm ls}$, and $D_{\rm l}$ represent the angular-diameter distances from the observer to the source, from the lens to the source, and from the observer to the lens, respectively. Note that the distances are normalized to values at the lens redshift $z_{\rm l}=0.5$ and the source redshift $z_{\rm s}=1$.

\begin{figure*}
   \centering
     \includegraphics[width=0.9\linewidth]{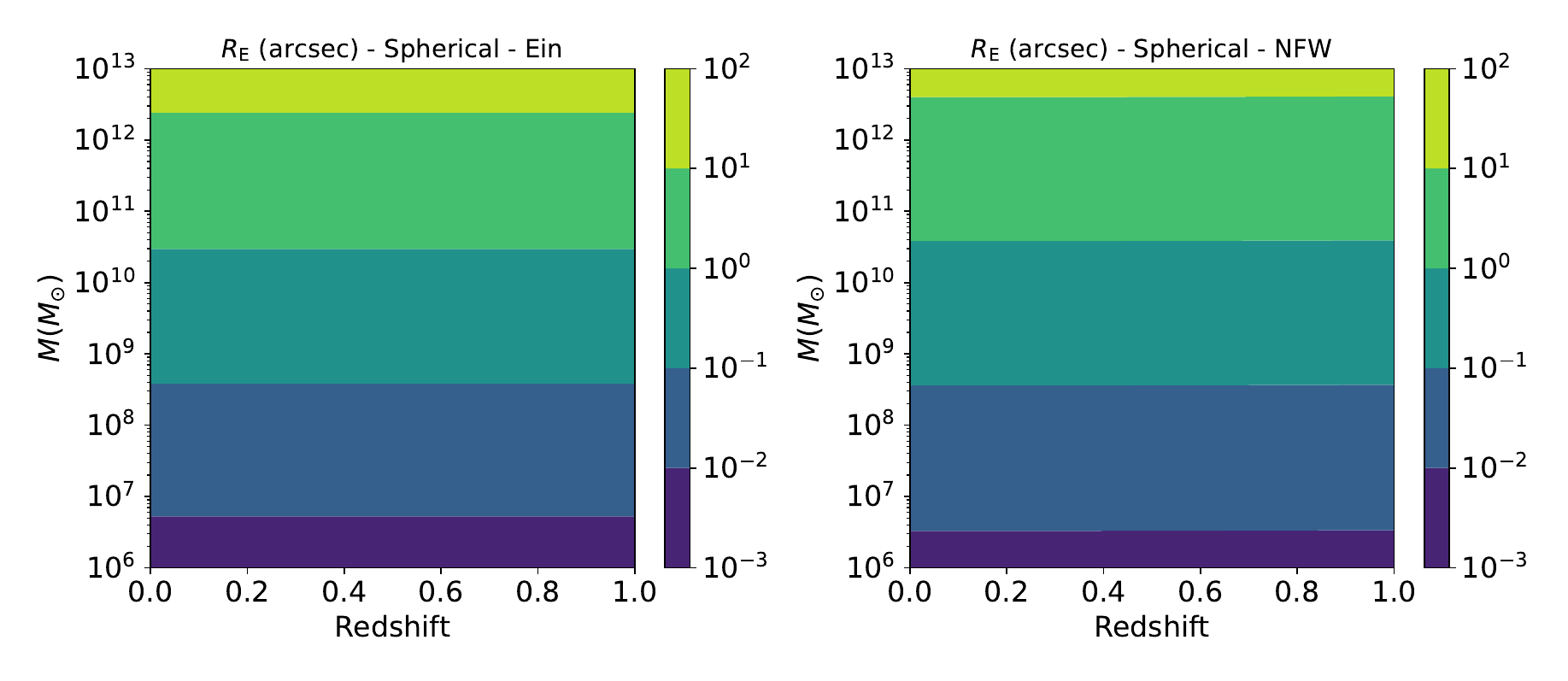}
     \includegraphics[width=0.9\linewidth]{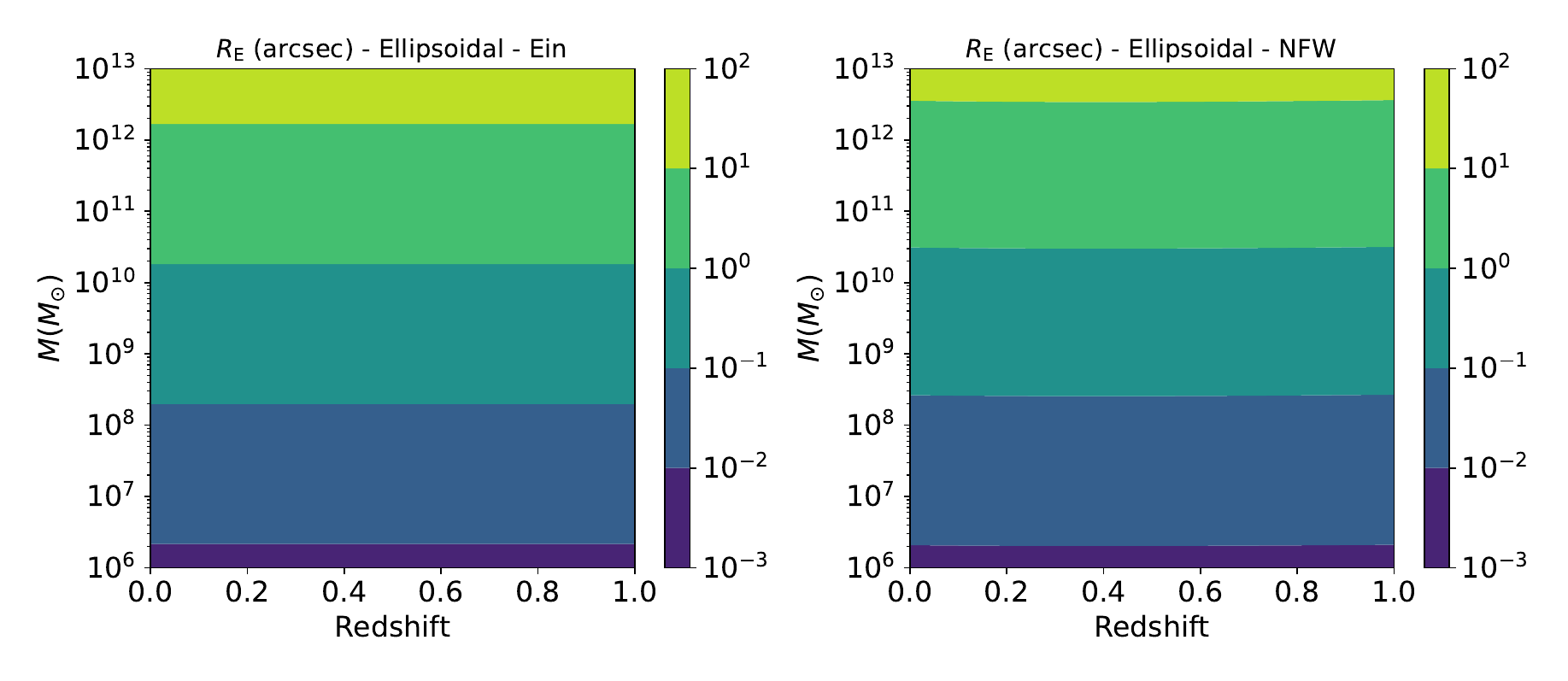}
     \includegraphics[width=0.9\linewidth]{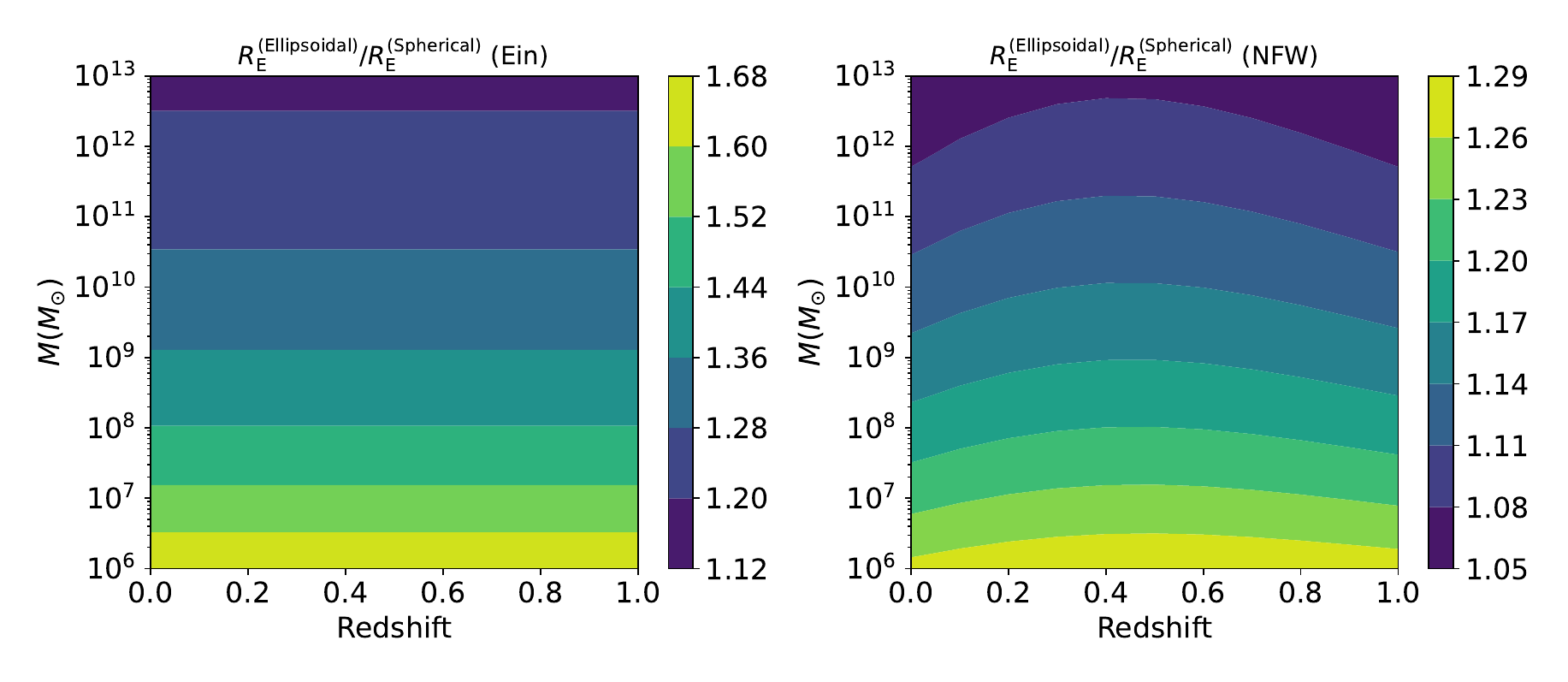}
   \caption{Einstein radii for gravitational lensing of GWs by ellipsoidal and spherical-collapse dark matter halos as a function of halo masses and cosmological redshifts. Top: Spherical-collapse dark matter halos with Einasto (left) and NFW (right) density profiles. Middle: Ellipsoidal-collapse dark matter halos with Einasto (left) and NFW (right) profiles. Bottom: Ratio of the Einstein radii for the ellipsoidal-collapse dark matter halo models to those obtained from the spherical-collapse dark matter halo models for Einasto (left) and NFW (right) profiles.}
   \label{Fig1}
\end{figure*}

In addition to point mass lenses, which consider compact objects like black holes as lenses, there are other more realistic models where galaxies, star clusters, and dark matter halos serve as lensing mediums. These can be described by the singular isothermal sphere (SIS) lens model. In the SIS model, the surface density is characterized by the velocity dispersion $v$ as $\Sigma(\xi)=v^{2}/\xi$.  In contrast, the surface mass density for a point mass model is given by $\Sigma(\xi)=M_{\rm l}\delta^2$. One can define $M_{\rm lz}$ as the mass within the Einstein radius, given by $M_{\rm lz}=4\pi^2 v^4 (1+z_{\rm l})D_{\rm ls}/(D_{\rm l}D_{\rm s})$. Consequently, the angular Einstein radius in the SIS model takes the following form:
\begin{eqnarray}
    \theta_{\rm ES}(M)=\sqrt{\frac{16G\pi^2 v^4}{c^2}\frac{D_{\rm ls}^2}{D_{\rm l}^2D_{\rm s}^2}} \hspace*{2cm}\nonumber\\
    \approx 3 \times 10^{-5}\left(\frac{v}{1{\rm km/s}}\right)^{2}\left(\frac{D_{\rm ls}}{D_{\rm l}D_{\rm s}}\right) \rm arcsec.
\end{eqnarray}
In the above relation, $v$ is the velocity dispersion of particles within $r_{\rm vir}$ \cite{2016MNRAS.456.3068O}
\begin{eqnarray}
    v=v_{15}\left(\frac{M}{10^{15}M_{\odot}h^{-1}}\right)^{1/3} \rm km/s, 
\end{eqnarray}
where $v_{15}$ is a constant in the original ellipsoidal-collapse model. For the NFW profile, $v_{15}$ is
\begin{equation}
v_{15} \approx 241 \ln(y^2) + 295 \ln(y) + 1156\,\mathrm{km/s},
\end{equation}
and for the Einasto profile, it takes the following form
\begin{equation}
v_{15} \approx b_{1}(\alpha) \ln(y^3) + b_{2}(\alpha) \ln(y^2) + b_{3}(\alpha) \ln(y)+ b_{4}(\alpha)\,\mathrm{km/s}.
\end{equation}
In the above reltions,
\begin{equation}
y=\frac{0.42+0.20 \nu^{-1.23} \pm 0.083 \nu^{-0.6}}{(Ht)^{2/3}},
\end{equation}
where $H$ is the Hubble constant, $t$ represents the collapse time, and $b_{i}$'s are second-order polynomials of the shape parameter \footnote{In this work, $\alpha= 0.115+0.014 \nu^{2}$.}. For $0.1 < \alpha < 0.52$, the polynomials can be determined as follows \citep{2016MNRAS.456.3068O}
\begin{eqnarray}
b_{1}(\alpha) = -173\alpha^{2} + 237\alpha - 14, \hspace*{1cm} \nonumber \\~ \nonumber \\
b_{2}(\alpha) = -389\alpha^{2} -378\alpha +287, \hspace*{1cm} \nonumber \\~ \nonumber \\
b_{3}(\alpha) = 1540\alpha^{2} - 195\alpha +244, \hspace*{1cm} \nonumber \\~ \nonumber \\
b_{4}(\alpha) = 71\alpha^{2} -287\alpha +1205. \hspace*{1cm} 
\end{eqnarray}

\begin{figure*}
   \centering
     \includegraphics[width=0.9\linewidth]{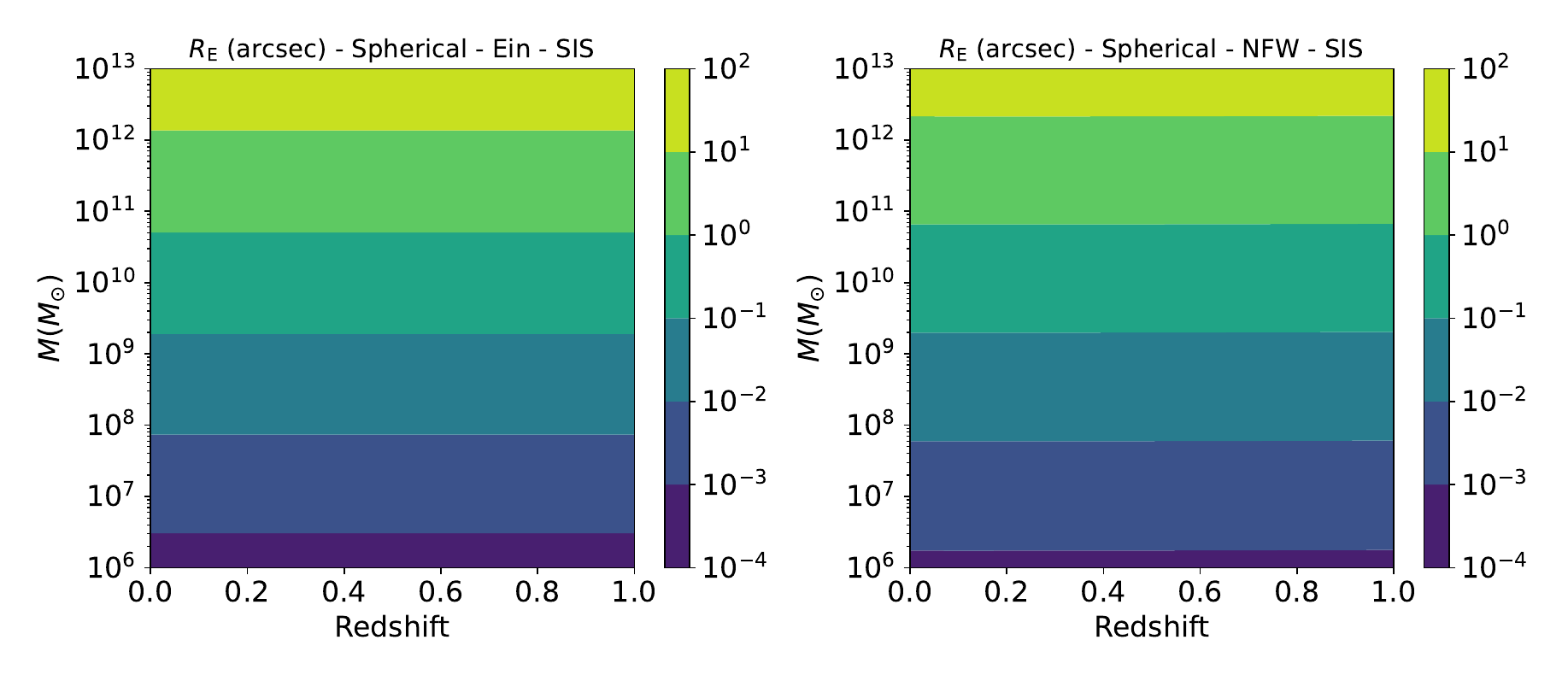}
     \includegraphics[width=0.9\linewidth]{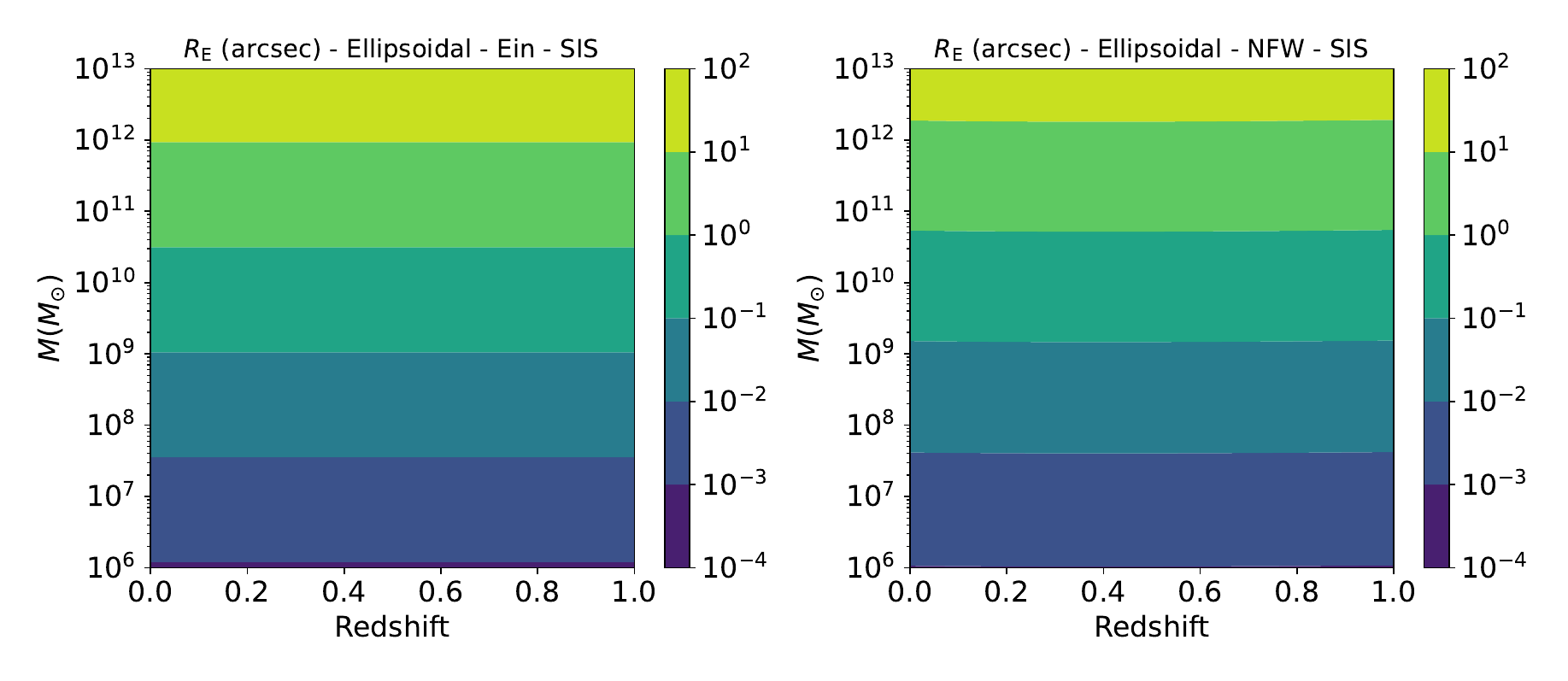}
     \includegraphics[width=0.9\linewidth]{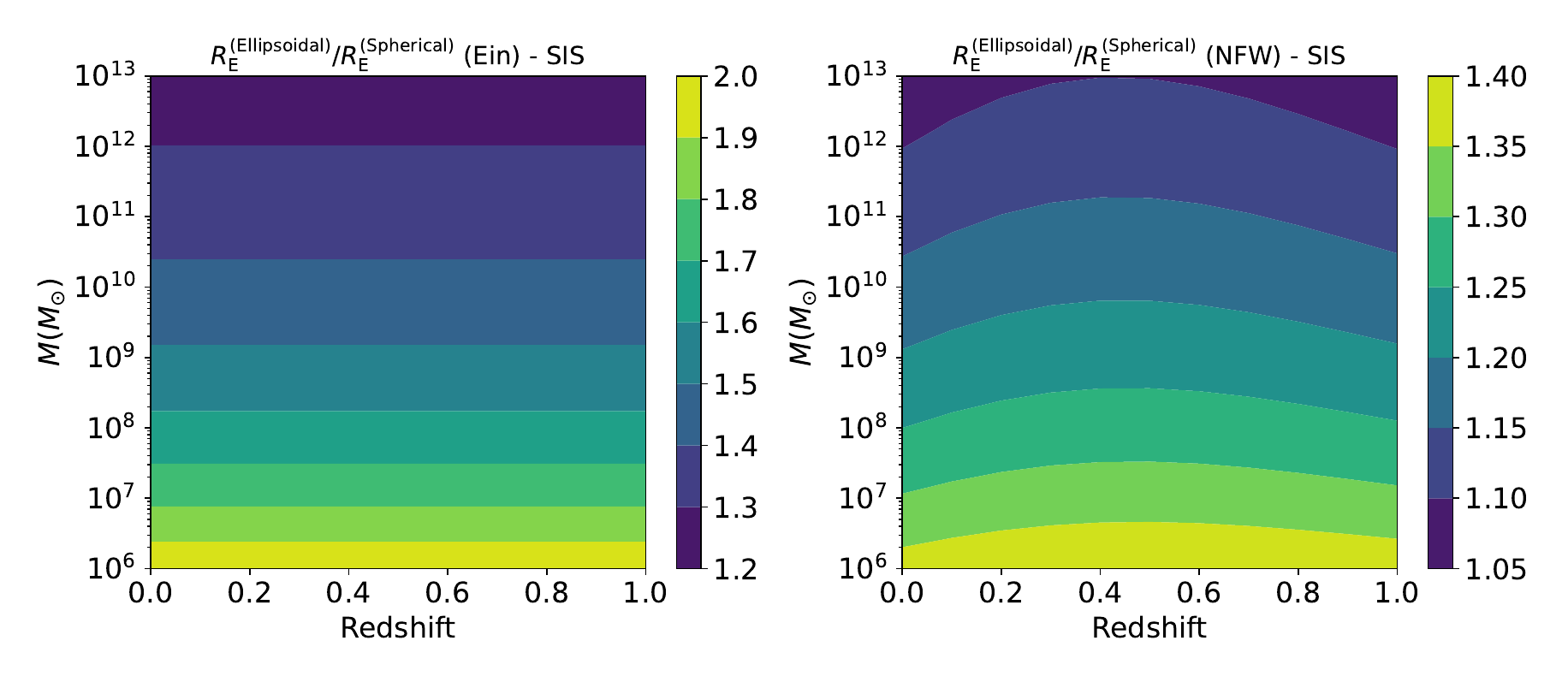}
   \caption{Similar to Fig.\,\ref{Fig1} but for the SIS model.}
   \label{Fig2}
\end{figure*}

In Fig.\,\ref{Fig1}, we have presented the Einstein radii for gravitational lensing of GWs across a range of halo masses $10^{6}\mbox{-}10^{13}M_{\odot}$ and cosmological redshifts for different dark matter halo models. The top panels present the Einstein radii for spherical-collapse dark matter halo models using the Einasto (left) and NFW (right) density profiles. The middle panels show the Einstein radii for ellipsoidal-collapse dark matter halo models with the same density profiles. The bottom panels display the ratio of the Einstein radii for ellipsoidal-collapse models to spherical-collapse models, for Einasto (left) and NFW (right) profiles. The Einstein radius is determined by assessing the concentration parameter for each halo model. As this parameter is linked to mass via the mass-concentration relationship, the Einstein radius can be computed based on the halo mass specific to each model.

As it is clear from the top panels, across the entire range of halo masses and redshifts, the Einasto profile yields Einstein radii that are consistently larger than those of the NFW profile by up to $5\mbox{-}30\%$. This advantage of the Einasto profile can be attributed to its more extended density distribution, which enhances the lensing effects compared to the more centrally concentrated NFW profile.

The middle panels indicale that the Einstein radii for ellipsoidal-collapse dark matter halo models are consistently larger than their spherical-collapse counterparts, with the difference becoming more pronounced at lower halo masses. For example, at the lowest halo masses considered, the Einstein radii for ellipsoidal-collapse dark matter halo models can be up to $50\mbox{-}68\%$ larger than those of spherical-collapse dark matter halo models. This substantial advantage arises from the more concentrated mass distributions of ellipsoidal halos compared to spherical halos, leading to stronger lensing effects.

As evident from the bottom panels, these ratios quantify the relative advantages of the ellipsoidal-collapse scenario over the spherical-collapse approximation. Across all halo masses and redshifts, the ratios for the Einasto profile range from approximately $1.12$ to $1.68$, while for the NFW profile they range from approximately $1.05$ to $1.29$. This indicates that the Einstein radii for ellipsoidal-collapse dark matter halo models are $5\mbox{-}68\%$ larger than those for spherical-collapse halos. Moreover, the ratios are generally higher for the Einasto profile compared to the NFW profile, with differences of up to $5\mbox{-}30\%$. This suggests that the Einasto profile better captures the enhanced lensing effects of ellipsoidal-collapse dark matter halos compared to the NFW profile.

We have also conducted similar calculations using the SIS model, with the results displayed in Fig.\,\ref{Fig2}. The panels of this figure reveal qualitatively similar outcomes to those obtained with the point mass model, underscoring the relative advantages of ellipsoidal-collapse models over spherical-collapse models and the superiority of the Einasto density profile over the NFW density profile for calculating Einstein radii. Additionally, the ratio of the Einstein radius from ellipsoidal-collapse models to that from spherical-collapse models shows that, for the Einasto density profile, the amplification ranges from $1.2$ to $2$, while for the NFW profile, it ranges from $1.05$ to $1.4$. The Einstein radii for ellipsoidal-collapse models are generally $20\mbox{-}100\%$ larger than those for spherical-collapse models, with the Einasto profile showing $15\mbox{-}40\%$ larger radii compared to the NFW profile. These differences underscore the importance of considering both the collapse model and the density profile when studying gravitational lensing effects.

In contrast to a point mass lens, an SIS lens, particularly at lower masses, possesses a smaller Einstein radius. This difference arises because a point mass can bend waves more effectively owing to its concentrated gravitational field, generally resulting in the formation of a single, highly magnified image.

The velocity distribution of dark matter particles within a halo can adhere to Maxwellian statistics. This means that the velocities of these particles are distributed according to a Maxwell-Boltzmann distribution, which describes the statistical behavior of particles in thermal equilibrium \cite{2012JPhCS.375a2048B}.
\begin{equation}
f(v)=\frac{4N}{\sqrt{\pi}v_{\odot}}\left(\frac{v/v_{\odot}}{1\rm km/s}\right)^{2}\exp\left(-\frac{v^{2}/v_{\odot}^2}{1\rm km/s}\right),
\end{equation}
where $v_{\odot}=220 $ km/s is the average velocity of particles in galactic halos at radius $r_{\odot}$, and $N=1.003$ is a normalizing constant. Such a distribution is characterized by a specific temperature and reflects how the velocities of the particles are spread out, with most particles having speeds around a certain average and fewer particles having very high or very low speeds.

This velocity distribution aligns with a simplified model often employed as a starting point: an isotropic sphere with uniform temperature and a density profile that scales inversely with the squared distance from the center, $\rho (r) \propto r^{-2}$. Although this model is advantageous due to its simplicity, it is likely insufficient to fully capture the true complexity of the halo's density and velocity distribution. Observations and numerical simulations indicate that dark matter halos deviate from this $1/r^2$ density profile, exhibiting more intricate structures. These complexities may include triaxial shapes and variations in velocity distribution along different axes.

\begin{figure}
   \centering
     \includegraphics[width=\linewidth]{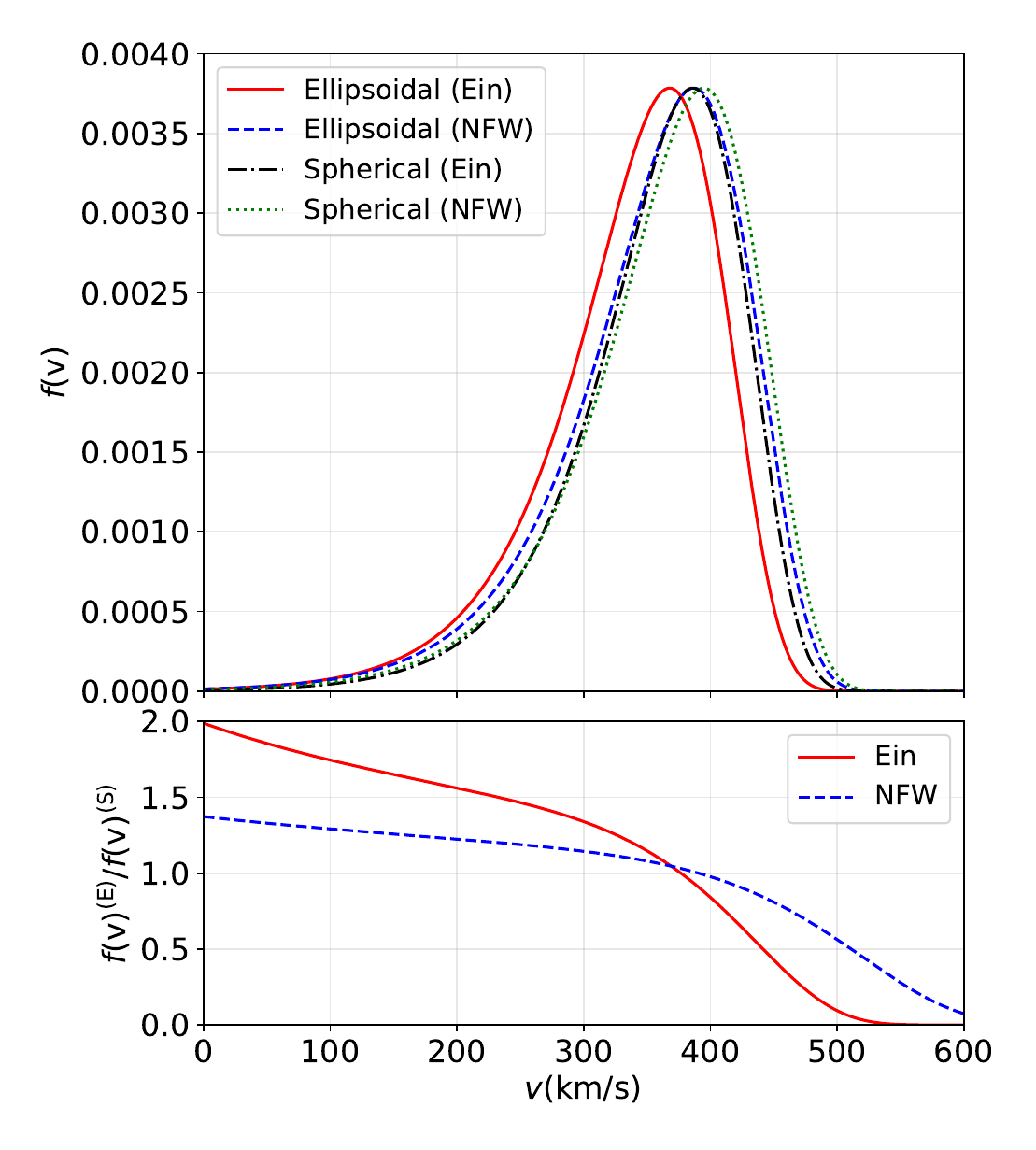}
   \caption{Top: The velocity distribution function of dark matter halos. This includes ellipsoidal-collapse models using Einasto (solid red line) and NFW (dashed blue line) density profiles, as well as spherical-collapse models with Einasto (dot-dashed black line) and NFW (dotted green line) density profiles. Bottom: The ratio of the velocity distribution function of dark matter halos, comparing ellipsoidal-collapse models to spherical-collapse models, for the Einasto (solid red line) and NFW (dashed blue line) density profiles.}
   \label{Fig3}
\end{figure}

In Fig.\,\ref{Fig3}, we have depicted the velocity distribution function for dark matter particles within different halo models. The top panel illustrates this function for both ellipsoidal and spherical-collapse dark matter halo models using the Einasto and NFW density profiles. The bottom panel depicts the ratio of the velocity distribution functions of the ellipsoidal-collapse models to the spherical-collapse models.
As shown in the figure, the velocity distribution curves vary across the different halo model and density profile combinations. These variations indicate that the shape of the halo and the collapse scenario influence the velocity distribution of dark matter particles. Specifically, the ellipsoidal-collapse models with both Einasto and NFW profiles exhibit higher peak values compared to their spherical-collapse counterparts. This suggests that ellipsoidal-collapse halos may contain a larger population of particles with higher characteristic velocities, which is supported by the bottom panel. It shows that the ratio of the velocity distributions for the ellipsoidal to spherical-collapse dark matter halos is generally greater than unity, particularly at typical velocities in galactic halos, i.e., $v \sim 200\mbox{-}300$ km/s. Additionally, differences between the Einasto and NFW profiles are evident. The Einasto profile shows a relatively higher velocity distribution at typical velocities in galactic halos compared to the NFW profile. However, the trend reverses at speeds exceeding $v \gtrsim 370$ km/s.

The probability of a source at redshift $z_{\rm s}$ being gravitationally lensed by an intervening mass distribution is given by a Poisson-like distribution as
\begin{equation}
 P = 1 - \exp\{-\tau(z_{\rm s})\},
\end{equation}
where $\tau(z_{\rm s})$ is the lensing optical depth, which represents the integrated surface mass density of potential lensing masses along the line of sight to the source.

To estimate the number of gravitationally lensed sources observed at a specific redshift, it is crucial to calculate the lensing optical depth, which depends on the distance to the sources. The lensing optical depth quantifies the cumulative probability of a source being lensed by the matter distribution along its line of sight. This includes the cross-sectional area and number density of potential lensing masses, such as galaxies and clusters, encountered by the light from the distant source to the observer \citep{2000ApJ...531..613B, 2006A&A...447..419F, 2023JCAP...07..007F}
\begin{equation}
\tau(z_{\rm s})=\int_0^{z_{\rm s}} \left|\frac{cdt}{dz}\right|dz \int \zeta\left(M_{\rm l},z\right) \frac{dn\left(M_{\rm l},z\right)}{dM_{\rm l}} dM_{\rm l},
\end{equation}
where $\zeta(M_{\rm l},z)$ represents the lensing cross-section, defined as follows:
\begin{equation}
\zeta(M_{\rm l},z)=\pi \theta_{\rm E}^2 D_{\rm l}^2,
\end{equation}
and
\begin{equation}
\frac{dt}{dz}=-\frac{1}{H(1+z)}.
\end{equation}
Also, ${dn(M_{\rm l},z)}/{dM_{\rm l}}$ represents the differential mass function, which quantifies the number density of dark matter halos categorized by their respective masses in various dark matter halo models (see Eqs.\,(\ref{massfunc} - \ref{nnuprime})).

\begin{figure}
\centering
\includegraphics[width=\linewidth]{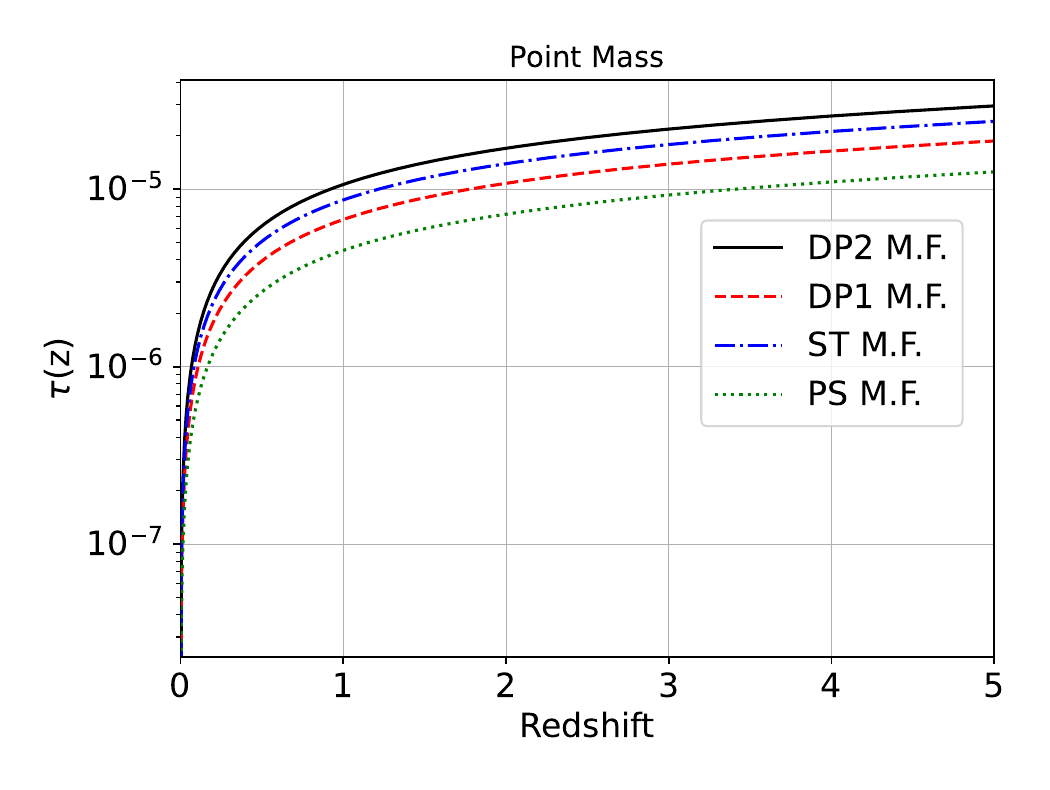} 
\includegraphics[width=\linewidth]{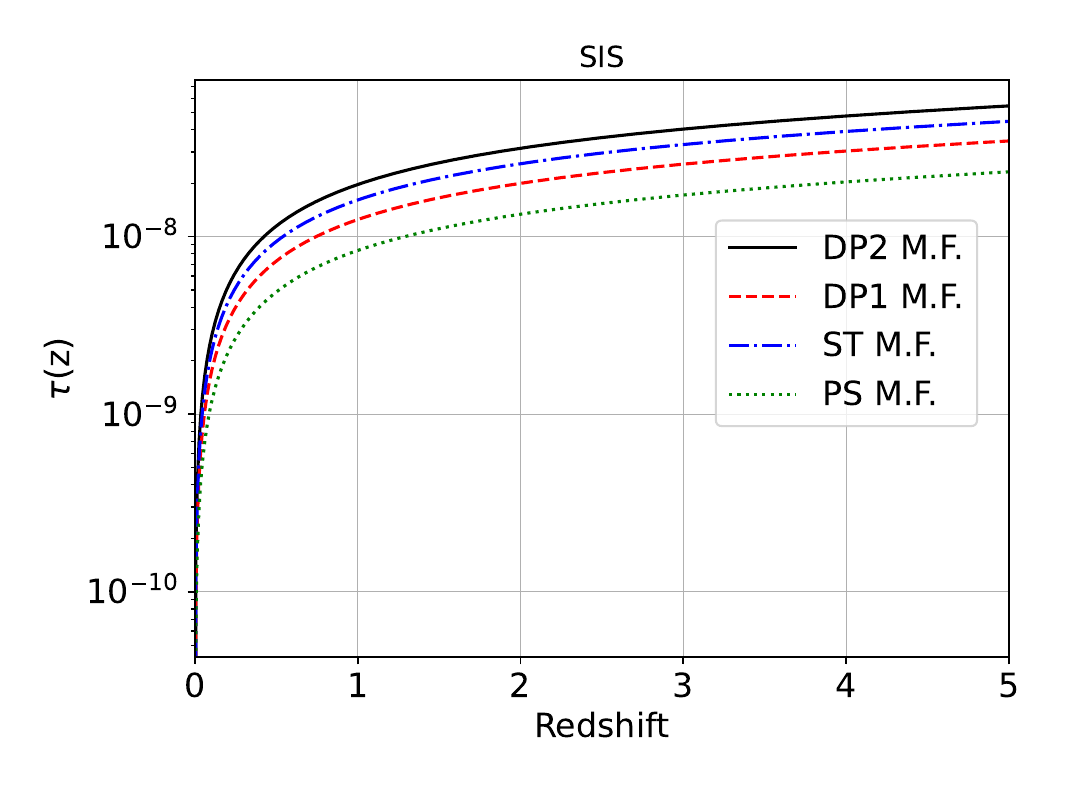} 
\caption{The lensing optical depth as a function of redshift, considering various mass functions: DP2 (solid black lines), DP1 (dashed red lines), ST (dot-dashed blue lines), and PS (dotted green lines). Top: For the point mass model. Bottom: For the SIS model.}
\label{Fig4} 
\end{figure}

In Fig.\,\ref{Fig4}, we show the lensing optical depth, which represents the integrated probability of a source being gravitationally lensed by intervening matter along the line of sight, as a function of redshift. The top panel considers the point mass model for lensing, while the bottom panel employs the SIS model. Both panels compare the lensing optical depth predicted by various mass functions, including the DP2, DP1, ST, and PS mass functions. The results indicate that the optical depth values in all models vary directly with redshift, which is reasonable given the expansion history of the Universe and the hierarchical nature of structure formation.

In the top panel, the DP2 mass function, which incorporates the effects of angular momentum, dynamical friction, and the cosmological constant, exhibits the highest lensing optical depth across the entire redshift range. This implies that the DP2 mass function predicts a higher probability of gravitational lensing events compared to the other mass functions considered. The ST mass function predicts a greater lensing optical depth compared to the DP1 mass function, which incorporates considerations of angular momentum and the cosmological constant. Furthermore, the optical depth predicted by the PS mass function is the lowest among all the examined mass functions. This underscores that the optical depth value for the simplified PS mass function is lower than those obtained from more realistic mass functions.

Similarly, in the bottom panel for the SIS model, the DP2, DP1, and ST mass functions yield higher lensing optical depths compared to the PS mass function. This trend is consistent with the point mass model in the top panel, indicating that the more realistic mass functions, which incorporate additional physical and geometrical factors, predict an increased likelihood of gravitational lensing events across various lens models. However, the optical depth values in this model are significantly lower than those obtained in the point mass model.
\begin{figure*}
   \centering
     \includegraphics[width=\linewidth]{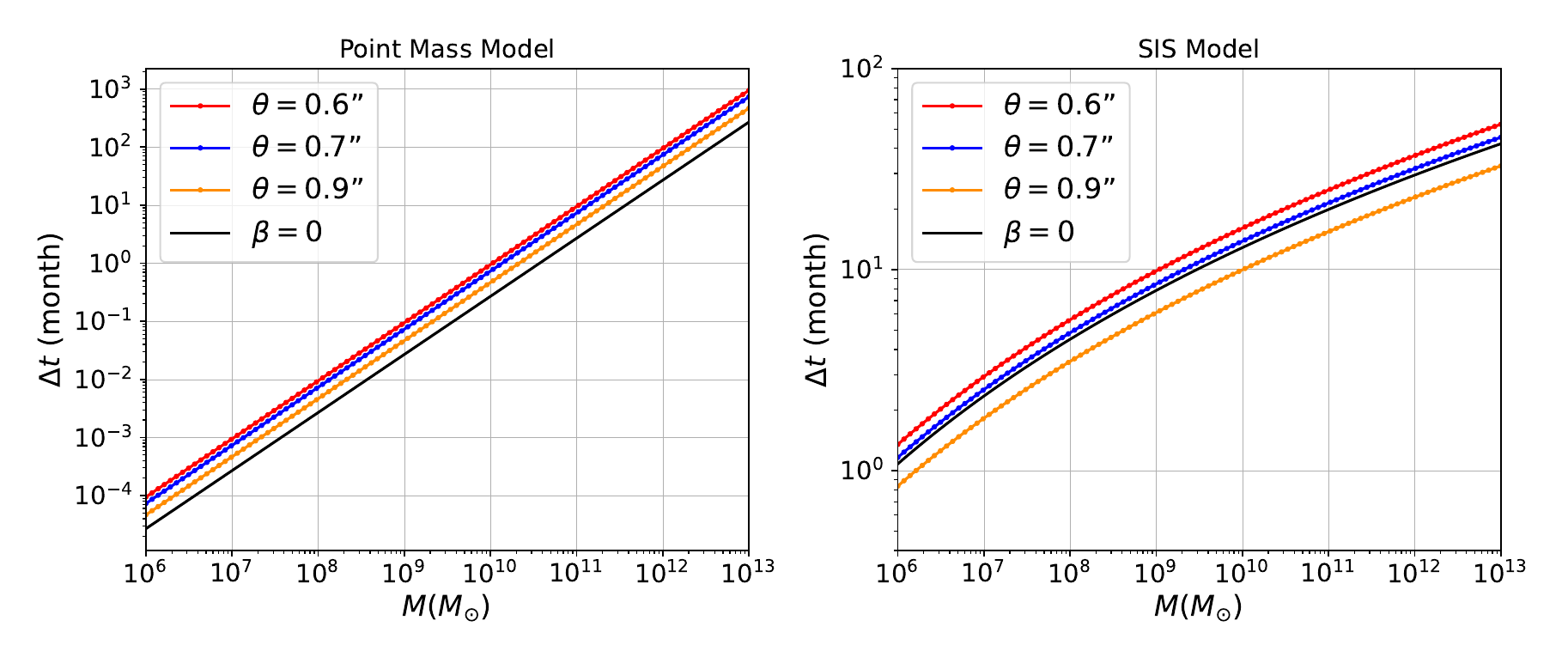}
   \caption{
The lensing time delay as a function of lens mass, taking into account the positions of the source and the image in the sky, for the point mass model (left) and the SIS model (right).}
   \label{Fig5}
\end{figure*}

When light from a source propagates through space, it can experience time delays in the presence of gravitational potential wells generated by massive objects. Strictly speaking, these potential wells are capable of deflecting the trajectory of light, which would otherwise travel in a straight line to the observer. Consequently, the observer detects the light waves at different times. We have considered two distinct cases: In the first case, there is a perfect alignment between the source, the lens, and the observer. In other words, the source lies precisely behind the lens along the observer's line of sight. In contrast, the second case assumes that the source subtends an angle with respect to the observer's line of sight, specifying its position relative to the lens.

We have adopted both the point mass and SIS lens models in gravitational lensing to investigate these two regimes profoundly. The lens equation relates the position of a source, $\beta$, and the position of the corresponding images, $\theta$, through the deflection angle, $\hat{\alpha}$ \footnote{For the first case of perfect alignment between the source, lens, and observer, the condition $\beta = 0$ is satisfied.}. In general, the small angle approximation and thin lens approximation can be considered. Then, the lens equation takes the following form \citep{2019RPPh...82l6901O}
\begin{equation}
    \beta = \theta - \frac{D_{\rm ls}}{D_{\rm s}} \hat{\alpha},
\end{equation}
where $\hat{\alpha}=4GM/c^{2}$. By utilizing the definition of the angular Einstein radius, one can obtain the following formula:
\begin{equation}
    \beta = \theta\left(1 - \frac{\theta_{\rm E}^2}{\theta^2}\right).
\end{equation}
This lens equation has two solutions, which represent the positions of the two lensed images formed by the gravitational lensing process
\begin{equation}
    \theta_{\pm} = \frac{\beta}{2} \pm \frac{\beta}{2} \sqrt{1+\frac{4\theta_{\rm E}^2}{\beta^2}},
\end{equation}
where positive and negative signs denote the first and second images, respectively. Typically, $\theta_{+}$ denotes the primary image, which is on the same side of the lens as the source, while $\theta_{-}$ refers to the secondary image, situated on the opposite side of the lens relative to the source.

As previously stated, the time delay in gravitational lensing refers to the difference between the travel time of a lensed, curved light ray forming one of the images and the travel time of an unlensed, straight light ray coming directly from the source. This time delay arises from two distinct contributions. The first contribution is the geometric delay, which results from the fact that the lensed light rays take different paths with varying orientations compared to the straight line path an unlensed ray would take. Since these curved trajectories have longer path lengths, it takes more time for the lensed rays to reach the observer, leading to a geometric delay. The second contribution is the gravitational delay, which is caused by the gravitational potential well of the lensing mass. As light rays pass through the curved spacetime around the lens, they experience a slowing down or time dilation effect predicted by general relativity. This gravitational time dilation adds an additional delay to the travel time of the lensed rays.

Under such conditions, one can derive distinct expressions for the point mass lens and the SIS lens models. Generally, the time delay equation is represented in the following form \cite{2019RPPh...82l6901O}
\begin{equation}
    \Delta{t_{\pm}} = \Delta{t_{\rm l}}\, \Phi (\theta_{+},\theta_{-}),
\end{equation}
where $\pm$ indices illustrate the time delay between two different images. $\Delta{t_{\rm l}}$ is the typical time delay for the lens which is,

Additionally, for the time delay between two different images, the $\pm$ indices are used. The typical time delay for the lens, $\Delta{t_{\rm l}}$ is given by
\begin{equation}\label{deltat}
    \Delta{t_{\rm l}} = (1+z_{\rm l}) \frac{4GM(<\theta_{\rm E})}{c^3},
\end{equation}
where $M(<\theta_{E})$ indicates that the Einstein radius acts as a probe, revealing the total projected mass of the lensing object enclosed within its radius. This is provided that the redshifts of both the lens and the source are known, and is expressed as
\begin{equation}\label{mtheta}
    M(<\theta_{\rm E}) = D_{\rm l}^{2} \int_0^{\theta_{\rm E}} 2\pi \theta^\prime \Sigma{(\theta^\prime)} \,d\theta^\prime = \pi D_{\rm l}^{2} \theta_{\rm E}^2 \Sigma_{\rm cr},
\end{equation}
where $\Sigma_{\rm cr}$ is the critical surface density, specified as
\begin{equation}\label{sigmacr}
    \Sigma_{\rm cr} = \frac{c^2}{4\pi G} \frac{D_{\rm s}}{D_{\rm l} D_{\rm ls}}.
\end{equation}
Using Eqs.\,(\ref{deltat}-\ref{sigmacr}), a clear description of the time delay of the lens can be obtained as follows
\begin{equation}
    \Delta t_{\rm l} = \frac{1+z_{\rm l}}{c} \frac{D_{\rm l}D_{\rm s}}{D_{\rm ls}} \theta_{\rm E}^2.
\end{equation}

Furthermore, the function $\Phi (\theta_{+},\theta_{-})$ is contingent upon the specified mass models outlined. In the case of a point mass lens, the expression is as follows \citep{2019RPPh...82l6901O}
\begin{equation}
    \Phi (\theta_{+},\theta_{-}) = \frac{2(\theta_{+}^{2} - \theta_{-}^{2})}{(\theta_{+} + \theta_{-})^2},
\end{equation}
and for the SIS model, it can be determined as
\begin{equation}
     \Phi (\theta_{+},\theta_{-}) = \frac{(\theta_{+}^{2} - \theta_{-}^{2})}{2\theta_{+}\theta_{-}} + \ln \left(\frac{\theta_{+}}{\theta_{-}}\right).
\end{equation}

The assessment of $\Delta{t_{l}}$ varies depending on the specific lens models applied. In the instance of a point mass lens, where there's a perfect alignment, meaning that $\left|\theta_{+}\right|$ is approximately equal to $\left|\theta_{-}\right|$, and $\beta = 0$, this results in $\Phi (\theta_{+},\theta_{-})\approx 0$. Consequently, when such conditions prevail, the multiple image arrangement exhibits symmetry; however, if this alignment is not perfect, the configuration becomes asymmetric.

In Fig.\,\ref{Fig5}, we have presented the lensing time delay as a function of the lens mass, taking into account the positions of the source and the image in the sky, for the point mass model and the SIS model.

The left panel illustrates the lensing time delay for the point mass lens model. The time delay is plotted against the lens mass, ranging from $10^{6} \mbox{-} 10^{13}\,M_{\odot}$. Multiple curves are shown, corresponding to different values of the source position parameter, which represents the angular separation between the source and the lens-observer line of sight. The time delay increases monotonically with increasing lens mass, as expected from the increased gravitational potential well of more massive lenses. Additionally, the time delay is minimum when $\beta=0$, which corresponds to the perfect alignment case where the source lies precisely behind the lens along the observer's line of sight. Also, as the source position $\beta$ increases from $0.6$ to $0.9$ arc seconds, the time delay decreases for a given lens mass. This behavior can be attributed to the decreasing asymmetry in the image configuration, leading to smaller differences in the path lengths traveled by the lensed images, thereby decreasing the time delay between them.

The right panel depicts the lensing time delay for the SIS lens model, which is generally considered a more realistic representation of gravitational lensing by extended matter distributions, such as galaxies or dark matter halos. Similar to the point mass model, the time delay increases with increasing lens mass, reflecting the stronger lensing effects of more massive systems. However, compared to the point mass model, the time delays for the SIS model are generally larger across the entire mass range. This can be attributed to the more complex gravitational potential of the SIS model, which induces greater deflections and path length differences for the lensed images. Furthermore, the influence of the source position $\beta$ on the time delay is also evident, with larger values of $\beta$ resulting in increased time delays due to the more asymmetric image configurations.

The results presented in this highlight the importance of considering both the lens mass and the relative positions of the source and images when studying gravitational lensing time delays. While the point mass model provides a simple approximation, the SIS model offers a more realistic representation of extended lensing systems, predicting larger time delays that may have significant implications for cosmological studies.
\section{Conclusions}\label{sec:iv}
In this study, we have explored gravitational lensing phenomena within the framework of more realistic dark matter halo models, moving beyond the over-simplified spherical-collapse dark matter halo models. Our analysis encompassed scenarios involving halo mass functions considering important geometrical and physical factors such as ellipsoidal-collapse condition, angular momentum, dynamical friction, and the cosmological constant. We have also incorporated the widely used density profiles of Einasto and NFW profiles into our gravitational lensing calculations and extended our analysis to two independent models of point mass lens and SIS lens. By opting for more realistic assumptions from dark matter halo models, we have determined crucial parameters associated with gravitational lensing, including Einstein radius, lensing optical depth, and the lensing time delay.

Our findings show that the Einasto density profile consistently yields larger Einstein radii compared to the NFW profile, attributed to its more extended density distribution. Ellipsoidal-collapse models also exhibit larger Einstein radii than spherical-collapse models, particularly at lower halo masses. The ratios of Einstein radii for ellipsoidal-collapse to spherical-collapse models emphasize the advantages of the former. These results highlight the importance of considering collapse models and density profiles in gravitational lensing studies.

We have also calculated velocity distribution functions for dark matter particles across various halo models and density profiles. Notably, ellipsoidal-collapse models consistently exhibit higher peak values compared to spherical-collapse models, suggesting a potentially larger population of high-velocity particles. This trend is supported by the ratio of velocity distributions favoring ellipsoidal-collapse halos, particularly at typical galactic halo velocities. Differences between the Einasto and NFW profiles are also evident, with the Einasto profile showing higher velocity distributions at typical velocities but reversing the trend at higher speeds. These findings highlight the significant influence of halo shape and density profile on dark matter particle velocities, crucial for understanding galactic dynamics and gravitational lensing.

We have illustrated the lensing optical depth as a function of redshift, employing both the point mass model and the SIS model. Across the depicted mass functions, including DP2, DP1, ST, and PS, optical depth values vary directly with redshift, reflecting the expansion of the Universe and the hierarchical structure formation. Notably, the DP2 mass function consistently predicts the highest optical depth, implying a greater likelihood of gravitational lensing events compared to other mass functions. Similarly, the ST and DP1 mass functions yield higher optical depths than the PS mass function, emphasizing the significance of considering additional physical and geometrical factors in realistic mass functions. This trend holds true across both lens models, suggesting an increased probability of gravitational lensing events with more comprehensive mass functions. However, optical depth values in the SIS model are notably lower than those in the point mass model, indicating model-dependent variations in predicted lensing probabilities.

Finally, we have calculated the lensing time delay as a function of lens mass for both the point mass and SIS models. The point mass model exhibits increasing time delays with lens mass, with minimum delay at perfect source-lens alignment and decreasing delay with increasing source position parameter. Conversely, the SIS model predicts generally larger delays across all mass ranges due to its more complex gravitational potential, influenced by both lens mass and source-lens positions. These findings emphasize the importance of considering both factors in gravitational lensing time delay studies, with the SIS model offering a more realistic portrayal of extended lensing systems and potentially significant implications for cosmological research.

This study offers appropriate insights into gravitational lensing within more realistic dark matter halo models, presenting avenues for further improvement and refinement. Future work could focus on incorporating more sophisticated halo density profiles and mass functions, considering complexities, which may lead to deviations from simplistic models. Exploring the impact of different dark matter models, such as warm or self-interacting dark matter, could provide new perspectives on the nature of dark matter itself. Additionally, incorporating modified gravity theories and considering the effects of clustering, evaporation, accretion, and merging of primordial black holes within dark matter halos could further refine our understanding of halo dynamics and gravitational lensing signatures. Furthermore, integrating physical factors like baryonic feedback processes, non-thermal pressure components, and environmental effects may offer a more comprehensive description of dark matter halo dynamics. Despite advancements, uncertainties persist in accurately determining halo density profiles and mass functions, modeling collapse processes, and accounting for lens and source redshift distributions and cosmological parameters. Addressing these uncertainties through improved observational constraints, advanced simulations, and refined theoretical models will be crucial for advancing our understanding of gravitational lensing within dark matter halos.



\bigskip
\bibliography{draft_ml}
\end{document}